\documentclass[aps,pra,reprint,amsmath,amssymb,superscriptaddress]{revtex4-2}

\usepackage[utf8]{inputenc}
\usepackage[T1]{fontenc}
\usepackage{graphicx}
\usepackage[colorlinks=true,urlcolor=blue,citecolor=blue,linkcolor=blue,bookmarks=false,pdfstartview={FitH}]{hyperref}

\usepackage{color,soul}

\begin{document}

\title{Strong electron-phonon coupling and enhanced phonon Gr\"uneisen parameters \\
in valence-fluctuating metal EuPd$_2$Si$_2$}

\author{Mai Ye}
\email{mai.ye@kit.edu}
\affiliation{Institute for Quantum Materials and Technologies, Karlsruhe Institute of Technology, 76021 Karlsruhe, Germany}
\author{Mark Joachim Graf von Westarp}
\affiliation{Institute for Quantum Materials and Technologies, Karlsruhe Institute of Technology, 76021 Karlsruhe, Germany}
\author{Sofia-Michaela Souliou}
\affiliation{Institute for Quantum Materials and Technologies, Karlsruhe Institute of Technology, 76021 Karlsruhe, Germany}
\author{Marius Peters}
\affiliation{Institute of Physics, Goethe-University Frankfurt, 60438 Frankfurt am Main, Germany}
\author{Robert M\"oller}
\affiliation{Institute of Physics, Goethe-University Frankfurt, 60438 Frankfurt am Main, Germany}
\author{Kristin Kliemt}
\affiliation{Institute of Physics, Goethe-University Frankfurt, 60438 Frankfurt am Main, Germany}
\author{Michael Merz}
\affiliation{Institute for Quantum Materials and Technologies, Karlsruhe Institute of Technology, 76021 Karlsruhe, Germany}
\affiliation{Karlsruhe Nano Micro Facility (KNMFi), Karlsruhe Institute of Technology, 76344 Eggenstein-Leopoldshafen, Germany}
\author{Rolf Heid}
\affiliation{Institute for Quantum Materials and Technologies, Karlsruhe Institute of Technology, 76021 Karlsruhe, Germany}
\author{Cornelius Krellner}
\affiliation{Institute of Physics, Goethe-University Frankfurt, 60438 Frankfurt am Main, Germany}
\author{Matthieu Le Tacon}
\email{matthieu.letacon@kit.edu}
\affiliation{Institute for Quantum Materials and Technologies, Karlsruhe Institute of Technology, 76021 Karlsruhe, Germany}

\date{\today}

\begin{abstract}
We study the valence crossover and strong electron-phonon coupling of EuPd$_2$Si$_2$ by polarization-resolved Raman spectroscopy. The fully-symmetric phonon mode shows strongly asymmetric lineshape at low temperature, indicating Fano-type interaction between this mode and a continuum of electron-hole excitations. Moreover, the frequency and linewidth of the phonon modes exhibit anomalies across the valence-crossover temperature, suggesting the coupling between valence fluctuations and lattice vibration. In particular, two phonon modes show significantly enhanced Gr\"uneisen parameter, possibly related to a nearby critical elastic regime.
The relative contribution of the structural change and valence change to the phonon anomalies is evaluated by density functional theory calculations.
\end{abstract}

\maketitle

\section{Introduction\label{sec:Intro}}

In the majority of $4f$-electron metals, the valence of the rare-earth element is essentially independent of temperature, because the compactness of $4f$ orbitals results in weak hybridization with the conduction bands. However, some rare-earth metallic systems, especially those with Ce, Sm, Eu, and Yb, exhibit a valence crossover in a narrow temperature range~\cite{Varma1976,Muller1982}.
The valence fluctuations in these metallic systems can lead to novel emerging physical properties. They could for instance induce effective interaction between itinerant electrons, leading to possibility of unconventional superconductivity. The superconducting state of CeCu$_2$Si$_2$ around 4.5\,GPa~\cite{Yuan2003,Seyfarth2012} and $\beta$-YbAlB$_4$~\cite{Naklatsuji2008} has been related to such valence fluctuations. On the other hand, 
 in Eu-based systems in which such valence-fluctuation-induced superconductivity remains elusive, a strong coupling to the lattice is seen.
 One of its most striking signature is certainly the dependence of the lattice volume to the valence state of the rare-earth. On cooling, electronic weight from the localized 4f shell is redistributed to itinerant conduction electrons, resulting in a spectacular contraction of the unit cell volume. 
One can naturally expect such strong dependence of the crystal structure to the local electronic one to strongly impact the dynamics of the lattice.

A particularly interesting case in this respect is that of intermediate-valence metal EuPd$_2$Si$_2$ which exhibits an Eu valence crossover from +2.25 at 300\,K to +2.75 at 20\,K, as determined from M\"ossbauer~\cite{Wortmann1985,Kemly1985,Scherzberg1985}, X-ray absorption~\cite{Wortmann1985,Kemly1985,Hisashi2011}, and photoemission~\cite{Martensson1982,Mimura2004a,Mimura2011} spectroscopies. Such a valence change is much larger than that in Ce- and Yb-based intermetallic systems, whose valence changes are typically only around 0.1~\cite{Kummer2018}. The valence crossover in EuPd$_2$Si$_2$ leads to an anomaly in the temperature dependence of magnetic susceptibility~\cite{Sampathkumaran1981} and specific heat~\cite{Wada2001}. The layered ThCr$_2$Si$_2$ (122) crystal structure of this system has a tetragonal crystal symmetry (point group D$_{4h}$; space group I4/mmm, No.139) which is preserved on cooling across the valence crossover but the difference of ionic radius for the larger Eu$^{2+}$ and the smaller Eu$^{3+}$, however, results in a volume decrease of the unit cell~\cite{Kliemt2022}.
Furthermore, it has been suggested that this system is close to a critical elasticity regime~\cite{Onuki2017}, akin to that recently proposed for the second-order critical endpoint of the pressure-induced Mott metal-insulator transition in organic compounds~\cite{Gati2016}. In such a regime, a critical enhancement of the Gr\"uneisen parameter results in a large lattice response to small changes in temperature and pressure~\cite{Zacharias2012}, and even though EuPd$_2$Si$_2$ is believed to be located on the high-pressure side of the critical endpoint~\cite{Onuki2017} one can expect significant fingerprints of this physics in the lattice dynamics of this compound.

As previously demonstrated for the case of Fe-based~\cite{Chauviere2009,Weilu2016,Shangfei2020} or Ni-based~\cite{Yao2022} superconductors, inelastic scattering of visible light (Raman scattering) is a perfectly suitable probe for measuring both electronic excitations and lattice dynamics in 122 crystal structures. It can therefore be used to gain some fresh insights regarding the electron-phonon coupling in EuPd$_2$Si$_2$. We note, however, that previous Raman works on Eu-based intermediate-valence metals~\cite{Zirngiebl1985,Zirngiebl1986,Cardona1991} focused on high-energy inter-multiplet excitations and that investigation of the Raman phononic response at low energies has not been reported so far.

In this work, we present polarization-resolved Raman spectra of EuPd$_2$Si$_2$  and first principle lattice dynamics calculations, which allow us to investigate the impact of valence crossover on the lattice dynamics in this system.
We observe a strongly asymmetric lineshape of the fully symmetric A$_{1g}$ phonon at low temperature as a result of Fano interference between this mode and a continuum of electronic excitations. Besides, we find that the frequency and linewidth of all four Raman-active phonon modes exhibit anomalous temperature dependence around the valence-crossover temperature.  This evidences a particularly strong electron-phonon interaction in EuPd$_2$Si$_2$ which constitutes a fingerprint of a nearby critical elastic regime and paves the way for systematic investigation of this phenomena.

\section{Method\label{sec:Exp}}

Single crystals of EuPd$_2$Si$_2$ were grown using the Czochralski technique as described in details in ref.~\cite{Kliemt2022}. Measurements of magnetic susceptibility indicated that the valence-crossover temperature T$_v$\,=\,120\,K, and X-ray diffraction shows that the volume of the unit cell decreases by 3.1\% on cooling~\cite{SM}.
Samples with cleaved $xy$ and $xz$ crystallographic plane were used. The sample surfaces were examined under polarized light to find a strain-free area. 

Raman-scattering experiments were performed with a Horiba Jobin-Yvon LabRAM HR evolution spectrometer. One notch filter and two Bragg filters were used in the collection optical path to clean the laser line from the back-scattered light. The samples were placed in a He-flow Konti cryostat. We used a He-Ne laser (632.8\,nm) with less than 1\,mW power that was focused on the sample with a x50 magnification objective. The laser spot size was around 5\,$\mu$m in diameter. 

Spectra were recorded with a 600\,mm$^{-1}$ grating and liquid-nitrogen-cooled CCD detector. The spectrometer resolution was 1.6\,cm$^{-1}$. All spectra of Raman response were corrected for the instrumental spectral response and Bose factor.

Five scattering geometries were employed to probe excitations in different symmetry channels and resolve all four Raman active phonons of the 122 structure. The relationship between the scattering geometries and the symmetry channels~\cite{Hayes1978} could be found in Table~\ref{tab:Exp}.

\begin{table}[b]
\caption{\label{tab:Exp}The relationship between the scattering geometries and the symmetry channels. For scattering geometry E$_{i}$E$_{s}$, E$_{i}$ and E$_{s}$ are the polarizations of incident and scattered light; X, Y, X', Y' and Z are the [100], [010], [110], [1$\overline{1}$0] and [001] crystallographic directions. A$_{1g}$, A$_{2g}$, B$_{1g}$, B$_{2g}$ and E$_{g}$ are the irreducible representations of the D$_{4h}$ group.}
\begin{ruledtabular}
\begin{tabular}{ll}
Scattering Geometry&Symmetry Channel\\
\hline
XX&A$_{1g}$+B$_{1g}$\\
XY&A$_{2g}$+B$_{2g}$\\
X'X'&A$_{1g}$+B$_{2g}$\\
X'Y'&A$_{2g}$+B$_{1g}$\\
XZ&E$_{g}$\\
\end{tabular}
\end{ruledtabular}
\end{table}

Density functional theory (DFT) calculations were performed using the mixed-basis pseudopotential method \cite{meyer97}. The exchange-correlation functional was represented by the generalized gradient approximation (GGA) \cite{perde96}. We applied the DFT\,+\,U approach with U=7\,eV for the Eu 4$f$ orbitals. We model the valence transition by considering two different electronic states: a ferromagnetic state which leads to a 4$f^{6.9}$ configuration (+2.1 Eu valence), and a nonmagnetic state having a 4$f^{6.1}$ occupancy (+2.9 Eu valence) \cite{SM}. We used the +2.1 valence state to approximate the high-temperature phase, and the +2.9 valence state to approximate the low-temperature phase. Phonon frequencies were calculated using the linear response or density-functional perturbation theory (DFPT) implemented in the mixed-basis pseudopotential method \cite{heid99}.

\section{Results and Discussion\label{sec:Res}}

\begin{figure}
\includegraphics[width=0.8\linewidth]{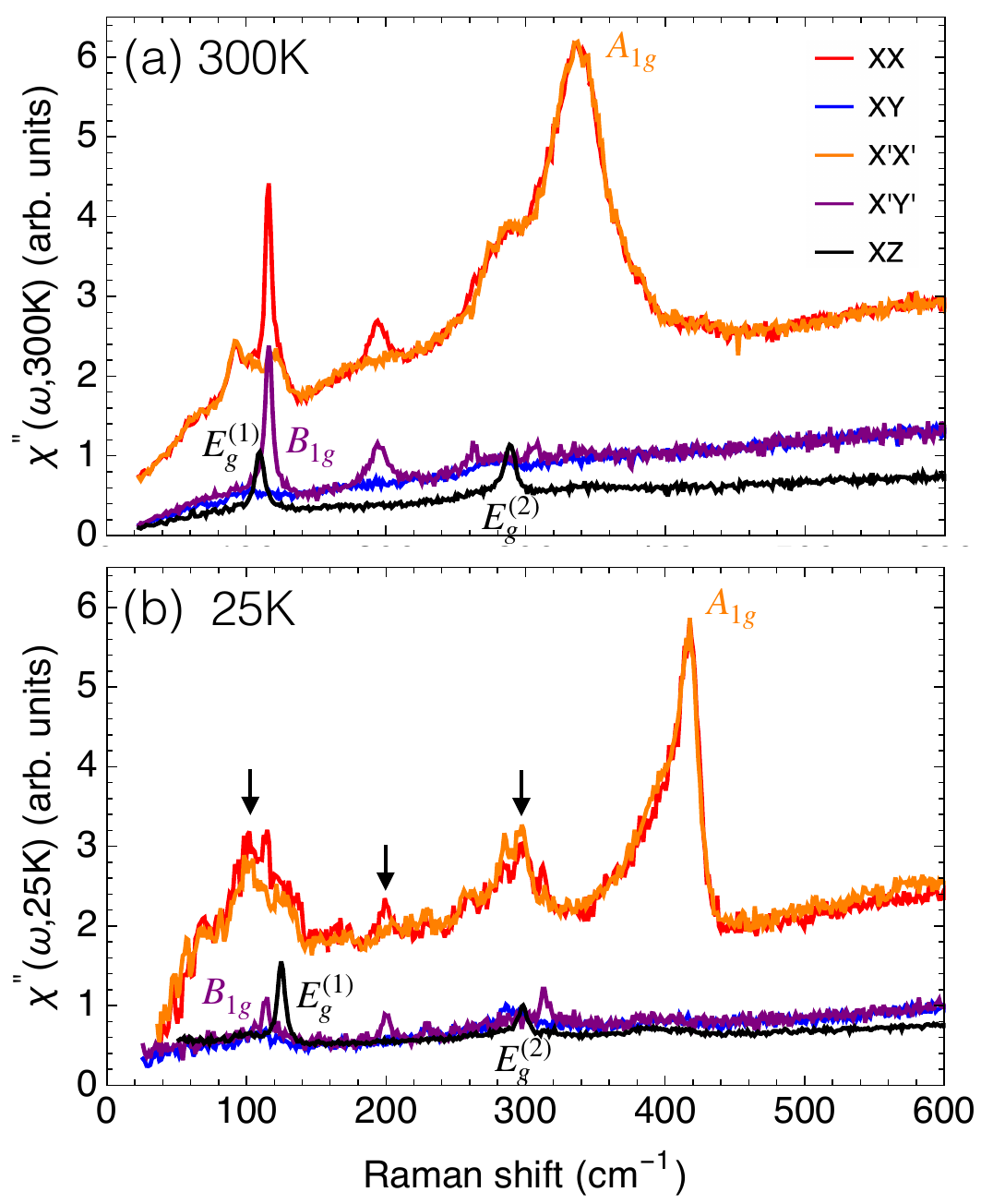}
\caption{\label{fig:Pola}Polarization dependence of the Raman spectra measured at (a) 300\,K and (b) 25\,K. The four Raman-active optical phonon modes are labelled by their respective symmetry. The three additional spectral features at around 110, 200, and 290\,cm$^{-1}$ are labelled by black arrows in (b)~\cite{SM}.}
\end{figure}

\begin{figure}[b]
\includegraphics[width=0.99\linewidth]{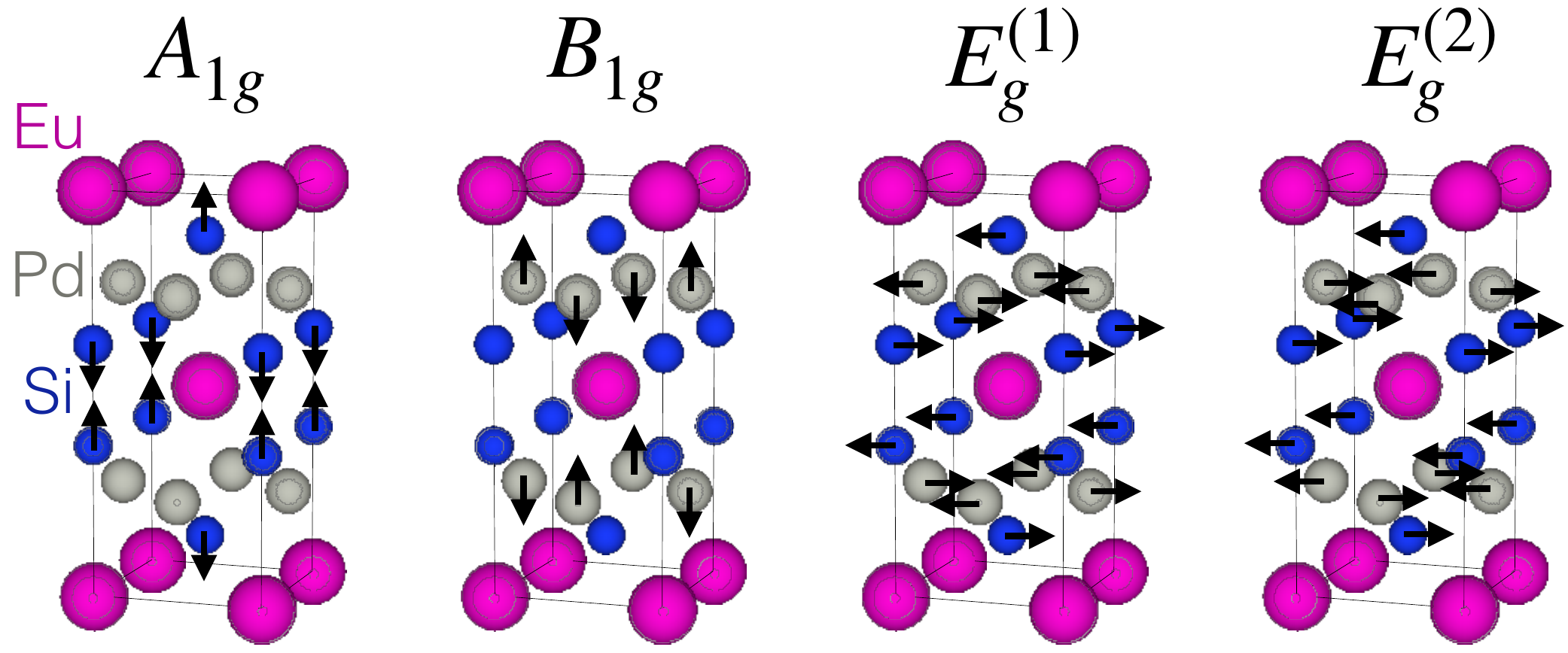}
\caption{\label{fig:Structure}The tetragonal crystal structure, and schematic vibration patterns of the Raman-active phonon modes in EuPd$_2$Si$_2$. These modes are classified by the irreducible representations of the tetragonal D$_{4h}$ point group.}
\end{figure}

In Fig.~\ref{fig:Pola}, we show the Raman spectra measured in the various scattering geometries listed in Table.~\ref{tab:Exp} at 300\,K and 25\,K. 
From group theory, four Raman-active phonon modes are expected in EuPd$_2$Si$_2$ with I4/mmm space-group symmetry: $1A_{1g}\oplus 1B_{1g}\oplus 2E_{g}$. The eigendisplacements corresponding to these modes are represented in Fig.~\ref{fig:Structure}. The A$_{1g}$ and B$_{1g}$ modes correspond to out-of-plane vibration of Si and Pd ions, respectively; the two E$_{g}$ modes are related to in-plane vibration of Si and Pd ions. According to Table.~\ref{tab:Exp}, the A$_{1g}$ mode appears in both XX and X'X' scattering geometries; the B$_{1g}$ mode appears in XX and X'Y' geometries; the E$_{g}$ mode appears only in XZ geometry. 
The four phonon modes are labelled in the experimentally measured spectra [Fig.~\ref{fig:Pola}].  Three additional weakly temperature dependent features, indicated by the black arrows in Fig.~\ref{fig:Pola} (b), are observed at about 110, 200, and 290\,cm$^{-1}$ and discussed in the supplemental material~\cite{SM}.
In each scattering geometry, we observe a continuum extending at least up to 1200\,cm$^{-1}$ and which is reminiscent of the incoherent electron-hole Raman response found in many correlated electron systems~\cite{Devereaux2007}. The continuum is most intense in the fully symmetric A$_{1g}$ channel; such phenomenon is at odds with the theoretically expected effects of the backflow corrections of the Raman response~\cite{Devereaux2007}, but in line with experimental observation for many systems, encompassing cuprates~\cite{Devereaux2007}, Fe-based superconductors~\cite{Kretzschmar2016}, iridates~\cite{Gretarsson2017}, and a variety of f-electron systems such as CeB$_6$~\cite{Ye2019} and YbRu$_2$Ge$_2$~\cite{Ye2019b}, to cite a few.

\begin{figure}
\includegraphics[width=0.9\linewidth]{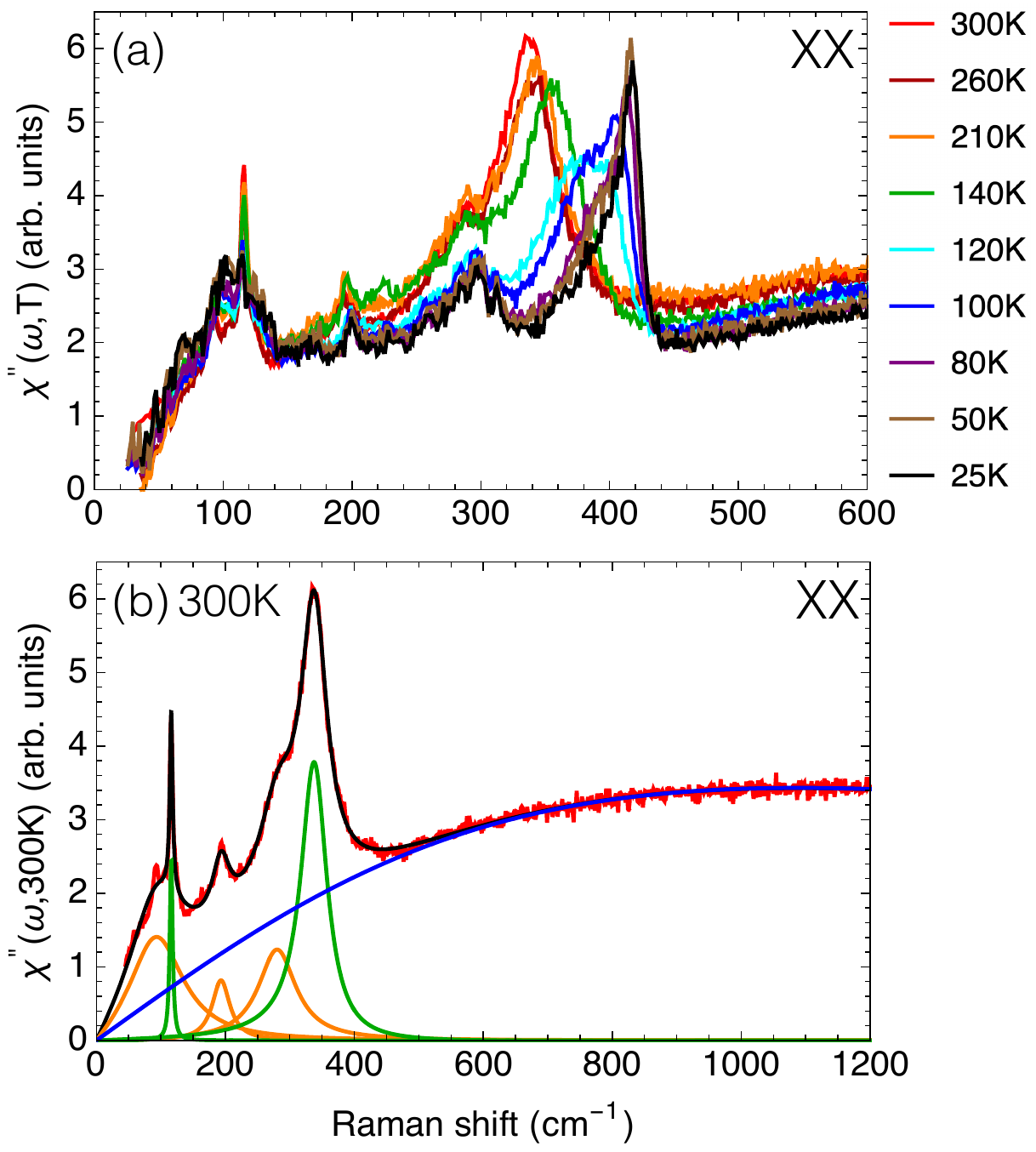}
\caption{\label{fig:Temp}Temperature dependence of the Raman spectra measured in the XX scattering geometry (a), and decomposition of the excitations in the XX geometry at 300\,K (b). In (b), the fitting curve is in black; the continuum of electronic excitations in blue; the B$_{1g}$ and A$_{1g}$ phonon modes in green, and the other modes in orange.}
\end{figure}

At 300\,K, the B$_{1g}$ and E$_{g}$ phonon modes show symmetric Lorentzian lineshape with a reasonably narrow linewidth of 5 to 10\,cm$^{-1}$, whereas the A$_{1g}$ mode is significantly broader. Sizeable changes in the spectra are observed upon cooling. The most striking changes are observed for A$_{1g}$ and B$_{1g}$ modes, in contrast to the E$_g$ phonons which seem to regularly harden and narrow.
At 25\,K,  the intensity of the B$_{1g}$ mode noticeably decreases; compared to the other phonon modes, the B$_{1g}$ mode also has other unique behaviors, which are discussed together later in this paper.
Meanwhile, the A$_{1g}$ mode exhibits strongly asymmetric lineshape, characteristic of Fano-type interference resulting from the interaction between this phonon mode and the underlying continuum of electronic excitations~\cite{Klein1975,Chen1993,Shangfei2020,Mai2021}. Moreover, in comparison to the other phonons, the hardening of the A$_{1g}$ mode upon cooling appears substantial. 
To gain more insight into the behavior of this mode, we show the temperature dependence of Raman spectra measured in the XX scattering geometry in Fig.~\ref{fig:Temp} (a). The A$_{1g}$ mode exhibits a continuous hardening on cooling, and it becomes more asymmetric at low temperature.

In order to quantify these changes, we use a Fano-interference model based on Green-function formalism~\cite{Klein1975, Chen1993} to fit the XX spectra, as illustrated in Fig.~\ref{fig:Temp} (b). The bare (non-interacting) phononic response for the A$_{1g}$ phonon mode has Lorenzian lineshape~\footnote{The function is symmetrized to ensure that the Raman response is zero at zero frequency and fulfill causality.}
\begin{equation}
G_{p}= -(\frac{1}{\omega-\omega_{p}+i\gamma_{p}}-\frac{1}{\omega+\omega_{p}+i\gamma_{p}})~,
\label{eq:fanoP}
\end{equation}
in which the parameters $\omega_{p}$ and $\gamma_{p}$ correspond to the mode frequency and half width at half maximum (HWHM), respectively.
These include temperature-dependent anharmonic effects. For the bare electronic response, we assume a relaxational form
\begin{equation}
G_{e}=\frac{1}{\omega_{e}-i\omega}~,
\label{eq:fanoE}
\end{equation}
with $\omega_{e}$ corresponding to the energy at which the electronic continuum displays its maximum intensity.
The response of the coupled phononic and electronic excitations can be obtained by solving the Dyson equation:
\begin{equation}
G=(G_0^{-1}-V)^{-1}.
\label{eq:fanoG}
\end{equation}
In Eq.~(\ref{eq:fanoG}),
\begin{equation}
G_0=\begin{pmatrix}G_{p} & 0\\
0 & G_{e}\end{pmatrix}~
\label{eq:fanoG0}
\end{equation}
is the bare Green's function and 
\begin{equation}
V=\begin{pmatrix} 0 & v\\
v & 0\end{pmatrix}~
\label{eq:fanoV}
\end{equation}
represents the coupling strength. As a result of this coupling, the spectral lineshape of the phonon is shifted and asymmetrically broadened, which can be quantified by the above given parameters~\cite{Chen1993,SM}. We further note that the parameter $v$ is related to, but distinct from the conventional electron-phonon coupling constant $\lambda$ defined as the ratio of the energy transferred from the lattice to the electrons to the change in lattice potential energy per atom~\cite{Ashcroft1976}. In particular, the coupling strength $v$ involves electron-phonon matrix elements which depends on the electronic structure.

The coupled Green function is converted to the experimentally measured Raman response by the following formula:
\begin{equation}
\chi^{\prime\prime}\sim\Im T^{T}GT~,
\label{eq:fanoChi}
\end{equation}
where $T^{T}=(\begin{array}{cc}t_{p}&t_{e}\end{array})$ represents the vertices for light scattering process (the superscript "T" denotes "Transpose"). 
The remaining spectral features observed in the XX scattering geometry are fitted by Lorentzian lineshapes~\cite{SM}. Despite of its simplicity, this approach provides good fits for the coupled electronic continuum and A$_{1g}$ phonon over a large frequency range at various temperatures with only six parameters in total~\footnote{The model does not fit the spectra near T$_v$ well enough compared to the spectra at other temperatures, possibly because of the strong fluctuations which are not accounted for in this model~\cite{Mai2021}.}. In particular, except for the scaling factor of the measured intensity, $t_{e}$, the electronic continuum has only one fitting parameter, $\omega_{e}$.

\begin{figure}
\includegraphics[width=0.99\linewidth]{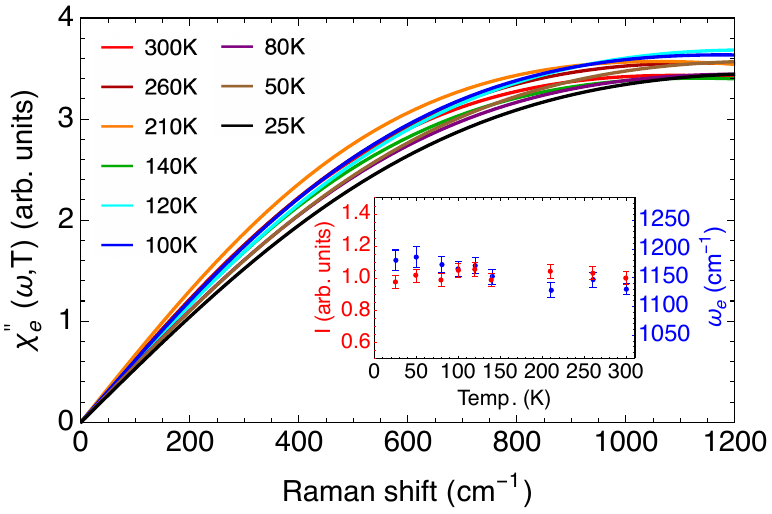}
\caption{\label{fig:Continuum}Temperature dependence of the continuum of electron-hole excitations. The left axis of the inset shows the temperature dependence for the integral, labelled by $I$, of experimentally measured Raman response between 800 and 1200\,cm$^{-1}$, normalized by the integral at 300\,K; the right axis of the inset shows the parameter of the electronic continuum, labelled by $\omega_{e}$, as a function of temperature.}
\end{figure}

\begin{figure}
\includegraphics[width=0.99\linewidth]{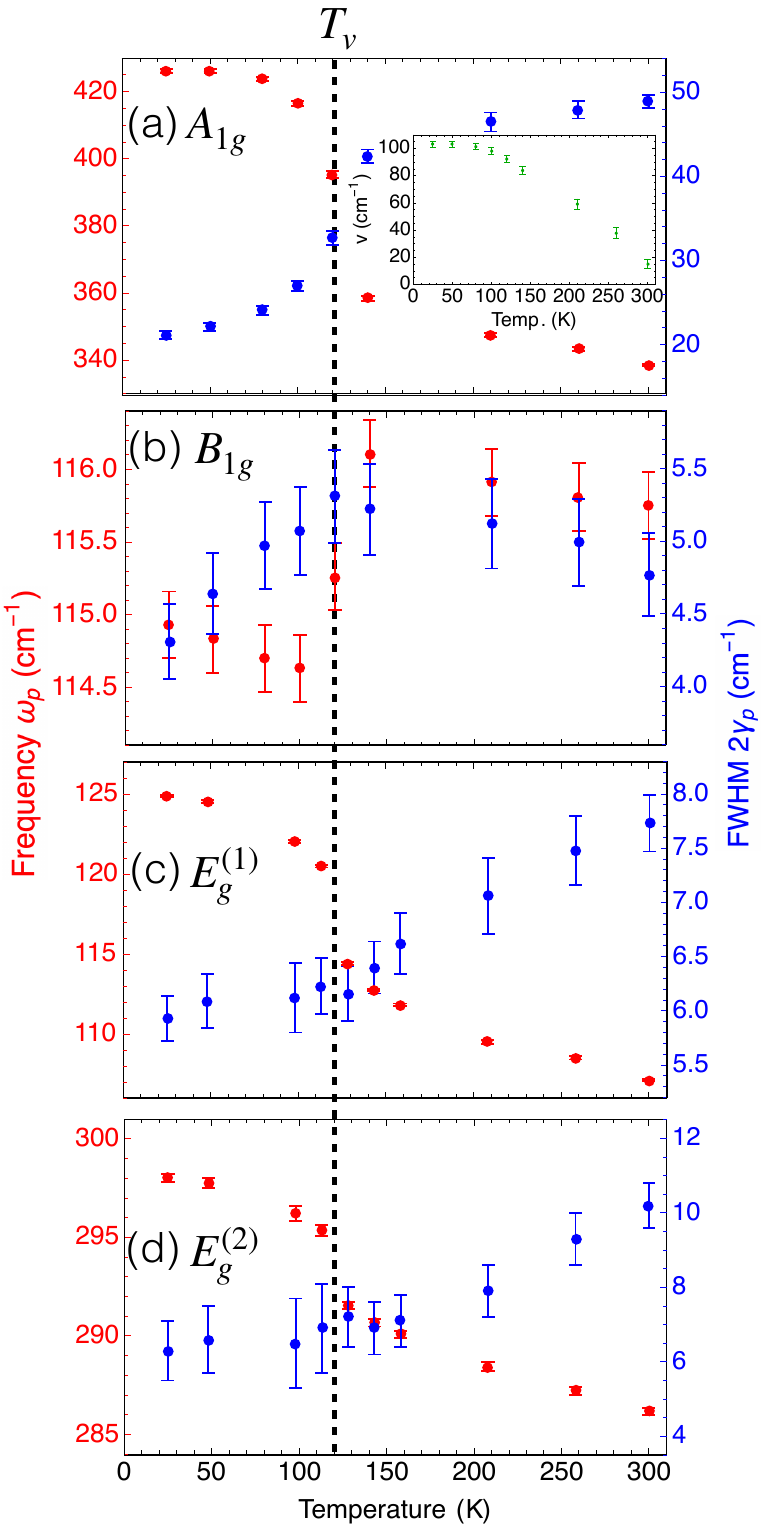}
\caption{\label{fig:Para}Temperature dependence of the frequency and full width at half maximum (FWHM) for the four Raman-active phonon modes of EuPd$_2$Si$_2$. The dashed line indicates the valence-crossover temperature T$_v$\,=\,120\,K. The coupling strength $v$ between the A$_{1g}$ phonon mode and electronic continuum is shown in the inset of panel (a).}
\end{figure}

In Fig.~\ref{fig:Continuum}, we present the continuum of electronic excitations $\chi^{\prime\prime}_e(\omega)=t_{e}^2\Im G_e~$, obtained from the fitting. As the magnitude of the continuum shows little temperature dependence [Fig.~\ref{fig:Continuum} Inset], the developing asymmetric lineshape of the A$_{1g}$ phonon mode on cooling must result mainly from an increasing coupling strength between this mode and the electron-hole excitations of the same symmetry. The phenomenological coupling strength parameter $v$ extracted from our analysis is shown in the inset of Fig.~\ref{fig:Para} (a) and indeed appear to increase on cooling. This effect is far from being trivial and relates to the temperature dependence of the itinerant electronic degrees of freedom across the valence transition.
The relevant electronic states are derived from Eu 5d, Pd 4d, and Si 3p conduction bands which are crossing the Fermi level~\cite{Song2022}. To gain further insights, a more elaborate theoretical model of the electronic response including explicit description of the electronic degrees of freedom and their fluctuations alongside proper treatment of vertex corrections (akin to seminal work performed for high-temperature superconductors~\cite{Devereaux1995,Bock1997}) might be needed but goes beyond the scope of the present work.

In Fig.~\ref{fig:Para}, we present the temperature dependence of the frequency $\omega_{p}$ and full width 2$\gamma_{p}$, obtained from our fitting, for the A$_{1g}$ mode. These are the bare parameters, which slightly differ from the apparent phonon frequency and linewidth shown in Fig.~\ref{fig:Temp}, with the latter ones renormalized by the Fano interaction.
We find that the relative hardening from 300\,K to 25\,K for the A$_{1g}$ and E$_{g}^{(1)}$ modes amounts to 24\% and 17\%, respectively. Such frequency change is much larger than the change observed in other Eu-based systems with a structural phase transition, for example EuFe$_2$As$_2$~\cite{Weilu2016}. Comparatively, the changes seen for the E$_{g}^{(2)}$ mode, with a frequency increase by 4\%, and the B$_{1g}$ mode with an anomalous frequency decrease by 1\%, are more modest but remain significant.
The fact that all phonons exhibit a marked change in their frequency just below T$_v$ indicates that these anomalous effects are related to the collapse of the unit cell volume, to the change of the Eu valency, or to a combination of both effects. 
The linewidth of the A$_{1g}$, E$_{g}^{(1)}$, and E$_{g}^{(2)}$ modes shows monotonic decrease on cooling, with the change of the linewidth for the A$_{1g}$ mode happening mostly around T$_v$. In contrast, the linewidth of the B$_{1g}$ mode exhibits an maximum near T$_v$. The latter anomaly further points to a strong electron-phonon coupling effect. 
The large frequency change for the the A$_{1g}$ and E$_{g}^{(1)}$ modes can be further quantified in terms of the phonon Gr\"uneisen parameter $\gamma_i = -(\Delta \omega_i/\omega_i)/(\Delta V/V)$. The volume changes are estimated on the basis of x-ray diffraction refinement measured at 300\,K and 80\,K ( the change of both phonon frequency and volume is small on cooling below 80\,K)~\cite{SM}. We obtain $\gamma_{A_{1g}}$\,$\approx$\,7.6 and $\gamma_{E_{g}^{(1)}}$\,$\approx$\,5.4; both are significantly larger than the typical values in metals ($\gamma$\,$\approx$\,2-3)~\cite{Ashcroft1976}. 
Each individual phonon contributes, weighted by its contribution to the heat capacity, to the total Gr\"uneisen parameter; it is therefore tempting to relate the unusually large value observed for these optical phonons to the previously suggested~\cite{Onuki2017} proximity to a critical elastic regime in which the Gr\"uneisen parameter is expected to diverge.


In Table~\ref{tab:DFT1}, we compare the experimentally measured phonon frequencies with the calculated ones. The DFT calculation captures the frequency increase for the A$_{1g}$, E$_{g}^{(1)}$, and E$_{g}^{(2)}$ modes, and the softening of the B$_{1g}$ mode.

\begin{table}
\caption{\label{tab:DFT1}Comparison between the experimentally measured frequencies and the calculated frequencies for the four Raman-active phonon modes. Two ground states were achieved by the DFT calculation; the one with +2 Eu valence was used to approximate the real system at 300\,K, and the one with +2.85 Eu valence was used to approximate the system at 25\,K. The unit for phonon frequencies is cm$^{-1}$.}
\begin{ruledtabular}
\begin{tabular}{ccccc}
&\multicolumn{2}{c}{300K}&\multicolumn{2}{c}{25K}\\
Mode&Exp.&Cal.&Exp.&Cal.\\
\hline
A$_{1g}$&339&315&419&416\\
B$_{1g}$&116&115&115&105\\
E$_{g}^{(1)}$&107&102&125&125\\
E$_{g}^{(2)}$&286&279&298&288\\
\end{tabular}
\end{ruledtabular}
\end{table}

From the DFT results, we gain further intuition into the relative contributions arising from the collapse of the unit-cell volume and the change of the Eu valence to the change of phonon frequencies. In Table~\ref{tab:DFT2}, we show how the phonon frequencies would change from their high-temperature value if there was only the volume effect or only the valence effect. The change of volume increases the frequency of all the modes, whereas the change of Eu valence mainly leads to a frequency decrease. Noticeably, the frequency change of the A$_{1g}$, E$_{g}^{(1)}$, and E$_{g}^{(2)}$ modes mainly results from the decrease of unit-cell volume, whereas the frequency change for the B$_{1g}$ mode almost entirely comes from the change of Eu valence. Such a significant difference might be associated to the unique behavior of the B$_{1g}$ mode, including the decrease of its intensity on cooling, the softening of its frequency, and the maximum of its linewidth appearing at T$_v$. We also note that the B$_{1g}$ mode is the only Raman-active phonon mode which does not involve the vibrations of Si atoms [Fig.~\ref{fig:Structure}].

\begin{table}
\caption{\label{tab:DFT2}The separate effect of volume decrease and valence increase on the phonon frequencies from DFT calculations. The frequencies corresponding to "HT Volume, HT Valence" are used to approximate the real system at 300\,K; these values are the same as the ones given in the third column of Table~\ref{tab:DFT1}. The short notations "HT" and "LT" stand for high-temperature and low-temperature, respectively. The unit for phonon frequencies is cm$^{-1}$; the frequency values in the brackets show the difference compared to the "HT Volume, HT Valence" case.}
\begin{ruledtabular}
\begin{tabular}{cccc}
&HT Volume&HT Volume&LT Volume\\
&HT Valence&LT Valence&HT Valence\\
Mode&&(Valence Effect)&(Volume Effect)\\
\hline
A$_{1g}$&315&327(+12)&388(+73)\\
B$_{1g}$&115&102(-13)&116(+1)\\
E$_{g}^{(1)}$&102&97(-5)&127(+25)\\
E$_{g}^{(2)}$&279&278(-1)&286(+7)\\
\end{tabular}
\end{ruledtabular}
\end{table}


For tetragonal Eu compounds EuTM$_2$X$_2$ (TM: transition metal; X: Si or Ge), Eu has around 2+ valence when the Eu-TM bond length is larger than 3.26\,\AA, whereas Eu has around 3+ valence when the Eu-TM bond length is smaller than 3.19\,\AA (See Fig.2 of Ref.~\cite{Song2022}). Only when the Eu-TM bond length is in between these two boundaries does a valence transition exist by varying temperature. In this respect, the fact that among the four Raman-active phonon modes, the B$_{1g}$ mode has the strongest dynamic modulation of the Eu-TM bond length, might explain why the frequency of the B$_{1g}$ mode is more strongly influenced by the Eu valence change.

The valence crossover near T$_v$ leads to an increase of the itinerant $f$ electrons, which make the system more metallic~\cite{Kliemt2022}. As the $f$ orbits are anisotropic, and considering the close relationship between the Eu valence and B$_{1g}$ mode, we suggest that the intensity decrease of the B$_{1g}$ mode on cooling, which mainly happens across T$_v$, could be related to the enhanced screening effect of the itinerant $f$ electrons.

\section{Conclusion\label{sec:Con}}

In summary, we have studied by inelastic light scattering the strong electron-phonon interaction in intermediate-valence metal EuPd$_2$Si$_2$.  The interaction between the A$_{1g}$ phonon mode and a continuum of electronic excitations of the same symmetry leads to Fano interference and strong asymmetric lineshape at low temperature.
We find anomalies of the phonon frequencies and linewidths near the valence-crossover temperature T$_v$, which evidence strong coupling between the valence fluctuations to the lattice dynamics. In particular, the A$_{1g}$ and E$_{g}^{(1)}$ phonon modes exhibit significantly enhanced Gr\"uneisen parameter, which possibly relates to the proximity to a critical elasticity regime. The frequency change of the A$_{1g}$, E$_{g}^{(1)}$, and E$_{g}^{(2)}$ modes mainly results from the volume collapse across T$_v$, whereas the softening of B$_{1g}$ mode frequency is probably mainly related to the Eu valence change.

Critical elasticity is commonly associated with instability in the acoustic channels. The experimental results for EuPd$_2$Si$_2$ presented here indicate optical phonons might also be dramatically renormalized approaching a critical elasticity regime. Further theoretical work is clearly needed to assess the generality of this interesting phenomena.

\begin{acknowledgments}
This work was funded by the Deutsche Forschungsgemeinschaft (DFG, German Research Foundation) - TRR 288 - 422213477 (projects A03 and B03).
S.M.S. acknowledges funding by the DFG – Projektnummer 441231589.
R.H. acknowledges support by the state of Baden-W\"{u}rttemberg through bwHPC. We thank Markus Garst and Roser Valenti for fruitful discussions.
\end{acknowledgments}


%

\end{document}